\documentclass[prl,twocolumn,amsmath,showpacs]{revtex4}
\usepackage{graphicx}

\begin{document}

\bibliographystyle{prsty}

\title {Magnetic ordering and spin waves in $\bf Na_{0.82}CoO_2$}

\author {S.~P. Bayrakci$^1$,
 I. Mirebeau$^2$,
 P. Bourges$^2$,
 Y. Sidis$^2$,
 M. Enderle$^3$,
 J. Mesot$^4$,
 D.~P. Chen$^1$,
 C.~T. Lin$^1$,
 and B. Keimer$^1$}

\affiliation{$^1$Max Planck Institute for Solid State Research,
Heisenbergstrasse 1, D-70569 Stuttgart, Germany\\
$^2$Laboratoire L\'eon Brillouin, C.E.A./C.N.R.S., F-91191
Gif-sur-Yvette CEDEX, France\\$^3$Institut Laue-Langevin, 156X,
38042 Grenoble Cedex 9, France\\$^4$Laboratory for Neutron
Scattering, ETH Zurich \& Paul Scherrer Institute, 5232 Villigen
PSI, Switzerland}\

\date{\today}

\begin{abstract}
Na$_x$CoO$_2$, the parent compound of the recently synthesized
superconductor Na$_x$CoO$_2$:$y$H$_2$O, exhibits bulk
antiferromagnetic order below $\sim\!20$ K for 0.75 $\leq x \leq$
0.9.  We have performed neutron scattering experiments in which we
observed Bragg reflections corresponding to A-type
antiferromagnetic order in a Na$_{0.82}$CoO$_2$ single crystal and
characterized the corresponding spin-wave dispersions. The spin
waves exhibit a strongly energy-dependent linewidth. The in-plane
and out-of-plane exchange constants resulting from a fit to a
nearest-neighbor Heisenberg model are similar in magnitude, which
is unexpected in view of the layered crystal structure of
Na$_{x}$CoO$_2$. Possible implications of these observations are
discussed.
\end{abstract}

 \maketitle

{The cobaltate Na$_{x}$CoO$_{2}:y$H$_{2}$O has recently enjoyed
intense attention. The composition with $x\sim 0.30$, $y\sim 1.4$
has been shown to be superconducting over a narrow range of $x$,
with maximum transition temperature $\rm T_c \sim 5$ K
\cite{takada03,schaak03}. This compound is particularly
interesting because its structure is similar to that of the
high-$\rm T_c$ copper oxide superconductors.  In both materials,
superconducting sheets containing oxygen and a spin-1/2 transition
metal are separated by layers of lower conductivity in an
anisotropic crystal structure.  However, a number of
characteristics suggest that the superconductivity in this
compound may be unusual in different ways from that found in the
cuprates.  For example, some experiments indicate that the
symmetry of the Cooper pair wavefunction may be $p$-wave
\cite{higemoto,fujimoto04}.

The unhydrated parent compound Na$ _{x}$CoO$_{2}$ is interesting
in its own right owing to its exceptionally high thermopower over
the range $ 0.5\leq x\leq 0.9$, which, unusually, accompanies low
resistivity and low thermal conductivity \cite{terasaki,mikami}.
Magnetic susceptibility data for compounds with $0.5 \leq x\leq
0.7$ show Curie-Weiss behavior with a negative Weiss temperature
\cite{ray,gavilano,wangNature}.
No static magnetic ordering has been observed for $x\leq 0.7$ in
$\mu$SR experiments \cite{sugiyama02}. The magnetism in this
system depends sensitively on the doping level: $\mu$SR
experiments performed on powder samples with $x=0.75$ suggested
the presence of a magnetically ordered phase below T = 22 K
\cite{sugiyama03}. Recent magnetic susceptibility measurements
have shown evidence of AF long-range order below 20 K for $0.75 <
x < 0.90$ \cite{spb04,mikami}.  Anisotropic DC magnetic
susceptibility and $\mu$SR measurements on the $x = 0.82$
composition both showed that the Co spins are oriented along the
{\it c}-axis \cite{spb04}.  Searches using unpolarized neutrons
for evidence of corresponding static magnetic order have thus far
been unsuccessful \cite{boothroyd04}.

In a recent time-of-flight experiment on a Na$_{x}$CoO$_{2}$
crystal with $x$ = 0.75, Boothroyd \textit{et al.} observed
ferromagnetic (FM) fluctuations within the $ab$-planes
\cite{boothroyd04}.  However, the $c$-axis momentum transfer could
not be varied independently of the energy transfer in this
experiment, and correspondingly only a projection of the
scattering cross-section onto the $ab$-plane was probed. The
magnetic ordering pattern below $\rm T_N$, as well as the exchange
parameters and their relationship to the macroscopic
susceptibility above $\rm T_N$, have thus far remained
undetermined. These issues are addressed in the
neutron scattering studies reported here.  For single crystals
with $x$=0.82, we find static ordering and low-energy fluctuations
characteristic of A-type antiferromagnetism: namely,
antiferromagnetically coupled ferromagnetic layers. Surprisingly,
the AF interlayer exchange constant resulting from a fit to a
simple Heisenberg model is almost as large as the intralayer FM
exchange, despite the difference in Co-Co distances. A possible
relationship of this observation to the Co$^{3+}$-Co$^{4+}$
charge-ordered states recently suggested on the basis of NMR
\cite{ning,mukhamedshin} and optical conductivity \cite{christian}
experiments is discussed.

Single crystals of $\gamma$-phase Na$_{x}$CoO$_{2}$ were grown by
the floating-zone technique in an image furnace \cite{spb04,chen}.
The sodium and cobalt contents were determined through analysis of
pieces cut from the same ingots and adjacent to the crystals used
for the neutron scattering experiments, using
inductively coupled plasma atomic emission spectroscopy (ICP-AES)
and atomic absorption spectroscopy (AAS). The Na:Co ratio was
found to be $x = 0.82\pm0.04$ for each of the two crystals
examined in this experiment.  Magnetic susceptibility measurements
were performed on portions of the ingots adjacent to the crystals
used in the experiment. All of the pieces tested exhibited the
expected AF transition at $\sim$20~K \cite{spb04}.

\begin{figure}
\includegraphics[width=1.00\linewidth]{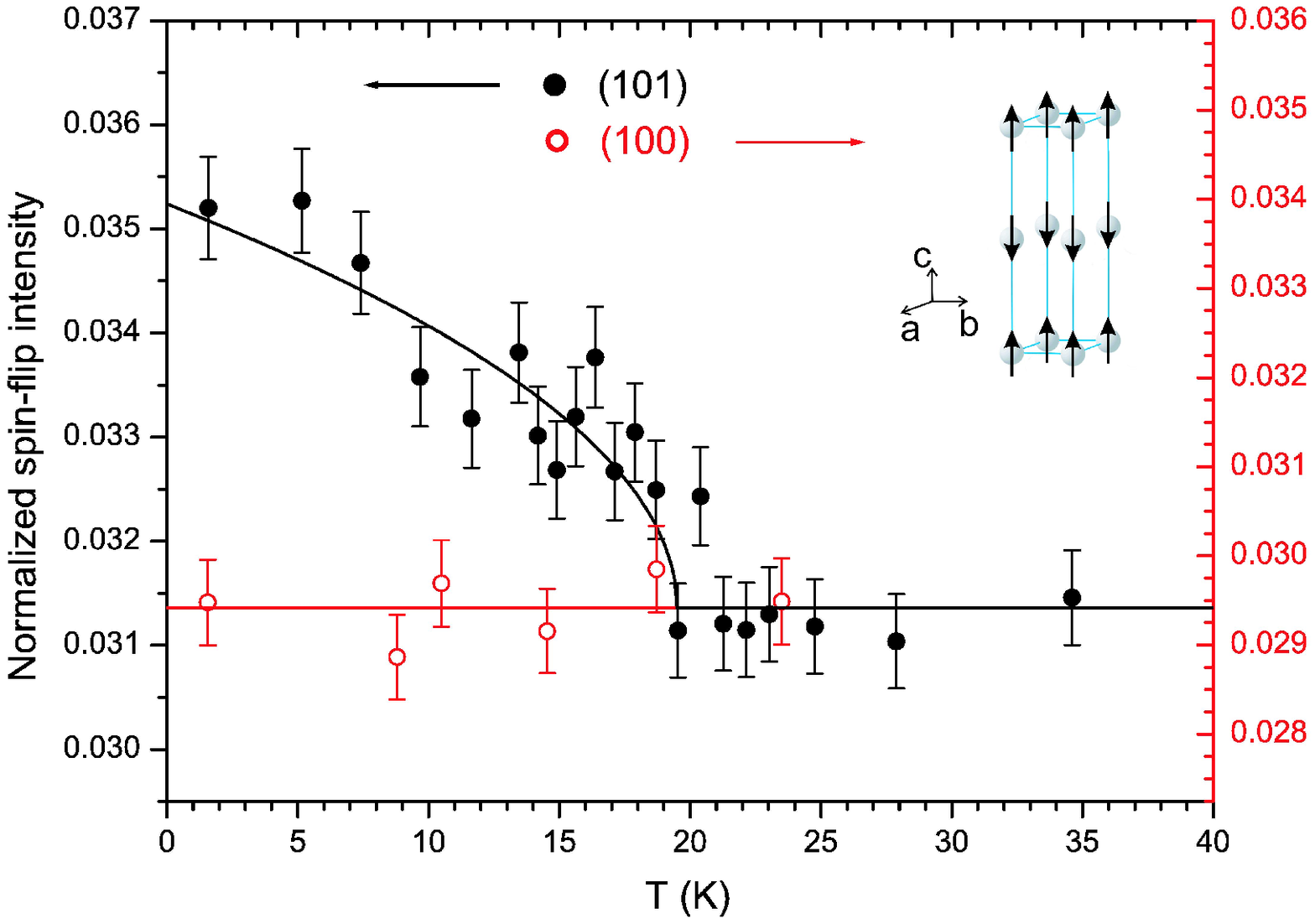}
\caption{Normalized SF intensity (equal to the SF intensity
divided by the NSF intensity, {\it viz.} the inverse of the
flipping ratio as it is usually defined; see text) at {\bf Q} =
(101) and (100), as a function of temperature. The data were taken
with the neutron polarization {\bf P} $\parallel$ {\bf Q}, with
$k_{i}$=2.662 $\rm \AA^{-1}$.  The lines are a guide for the eye.
Inset: the A-type AF structure, shown with Co spins $\parallel$
{\it c}.}
\end{figure}

We discuss first the determination of the magnetic ordering
pattern by elastic neutron scattering. We used the polarized-beam
spectrometer 4F1 at the Laboratoire L\'eon Brillouin, Saclay,
France to monitor the neutron scattering intensity in the
spin-flip (SF) and non-spin-flip (NSF) channels at a number of
high-symmetry points in the reciprocal lattice. Fig. 1 shows the
SF intensity (normalized to the NSF intensity) at the (100) and
(101) reciprocal-lattice vectors \cite{notation} as a function of
temperature, with the neutron polarization {\bf P} $\parallel$
{\bf Q}. Temperature-independent contributions to the intensity
are present at both positions.
These contributions originate mostly from the leakage of NSF
scattering from the corresponding nuclear Bragg reflections (which
are not forbidden at these reciprocal-lattice vectors), due to the
usual limitations of the instrument. For ${\bf Q} = (101)$, an
additional contribution to the SF intensity originating from
electronic magnetic scattering appears below $\rm T_N$. We also
observed magnetic intensity below $\rm T_N$ at (103), (105),
(111), and (113), whereas none was observed at (100) [see Fig. 1]
or (102), within experimental error. This indicates that the
magnetic propagation vector is (001). The unit cell of
$\gamma$-phase Na$_{x}$CoO$_{2}$ contains two planes of CoO$_{6}$
octahedra; a magnetic ordering vector of (001) therefore
corresponds to AF ordering in the {\it c}-direction, combined with
FM order within the $ab$-planes (Fig. 1, inset). The dependence of
the SF intensity on the neutron spin direction at the sample
confirms the inference from the uniform susceptibility of the
$c$-axis orientation of the magnetic moment \cite{spb04}. The
(001) Bragg reflection is unobservable because the spin
orientation factor in the elastic neutron scattering cross-section
vanishes for moments directed along {\it c}.  By comparing the
magnetic intensity at (101) with the nuclear intensity
\cite{lovesey} at (100),
using the isotropic form factor,
we extracted a value of $0.13\pm0.02\, \mu \rm _B$ per Co.

The inelastic experiment was performed on the spectrometer IN20 at
the Institut Laue-Langevin, Grenoble, France. We used an
unpolarized configuration with a double-variable-focusing Si(111)
monochromator and a PG(002) analyzer with fixed vertical and
horizontal focusing.  In order to maximize the flux for the
relatively small samples, which had volumes of 147 and 162~mm$^3$,
no collimation was employed. Most of the data were taken either
with a fixed final neutron wavevector of $k_{f}$=2.662 $\rm
\AA^{-1}$ or a fixed incident wavevector of $k_{i}$=4.1~$\rm
\AA^{-1}$.  The full width at half maximum (FWHM) of the energy
resolution at zero energy transfer is approximately 1 meV in the
former arrangement and 3 meV in the latter.  The samples were each
mounted in the (HHL) scattering plane and placed in a pumped
$^4$He cryostat.

\begin{figure}
\includegraphics[width=1.00\linewidth]{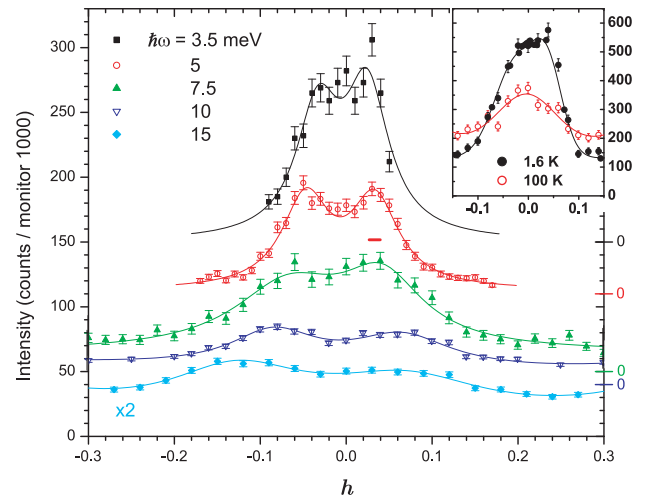}
\caption{Scans at constant energy transfer $\hbar\omega$ taken
along the $(hh0)$ direction through (001) (for $\hbar\omega\leq 5$
meV) or (003), at 1.6 K. The fits are to double Lorentzians. The
red bar indicates an estimate of the FWHM of the instrumental
resolution for the scan at $\hbar\omega=5$ meV. For clarity, the
data have been offset vertically for the scans with
$\hbar\omega<15$ meV (the corresponding vertical-axis zeroes are
indicated at right) and the 15 meV data have been scaled by a
factor of 2.  Inset: constant-energy scans along $(hh0)$ through
(001) at energy transfer of 5 meV, at 1.6 K and 100 K. The fits
are to double Gaussians. The axis units are the same as in the
main panel.}
\end{figure}


Low-temperature scans at constant energy transfer $\hbar\omega$
through the ordering wavevector (001) along the in-plane $(hh0)$
direction are shown in Fig.~2. Since the instrumental resolution
width is narrow relative to the peak widths, especially at large
energy transfers, the data shown were fitted to unconvoluted
Lorentzians. As a consequence of the large peak widths at higher
excitation energies, interference from an optical phonon at
$\hbar\omega = 20$ meV, and the lower neutron flux at larger
incident energies, it was not possible in this experiment to
determine the dispersion relation up to the in-plane magnetic zone
boundary.  The inset shows similar scans taken with energy
transfer $\hbar\omega$ = 5 meV at 1.6 K and 100 K, with coarser
instrumental resolution ($k_{i}=4.1 \rm \AA^{-1}$).
The integrated intensity
decreases with increasing temperature, characteristic of a
magnetic excitation. The separation between the two peaks
is indistinguishable from zero at 100 K, which suggests that
this signal arises from overdamped low-dimensional fluctuations.
Comparison of this low-temperature scan through (001) with an
equivalent one through (003) (not shown) demonstrates that the
intensity decreases with increasing $|\rm {\bf Q}|$, as expected
for a magnetic excitation. No magnetic intensity was detected in
similar scans through (002) and (004).

\begin{figure}
\includegraphics[width=1.00\linewidth]{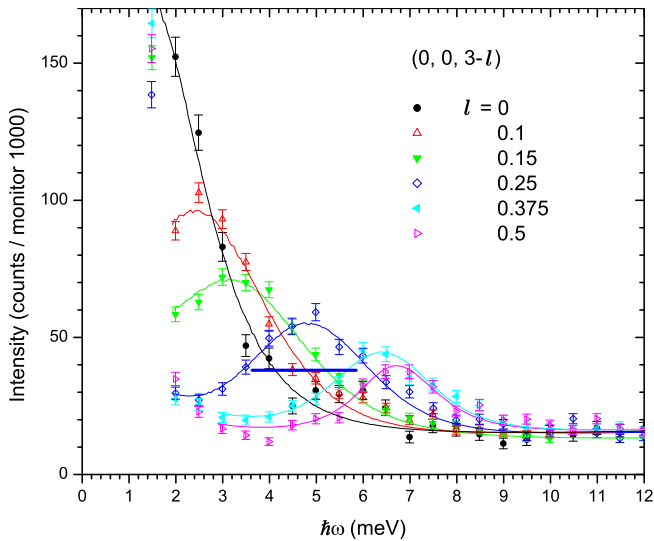}
\caption{Constant-{\bf Q} scans taken at different values of
(0,~0,~$3-l$), with T = 1.6 K and $k_{f}\!=\!2.662 \rm \AA^{-1}$.
The curves are the results of convolutions of the full spin-wave
dispersion (see Fig. 4) with the instrumental resolution
ellipsoid; the blue bar indicates an estimate of the FWHM of the
instrumental resolution for the scan with $l=0.25$.}
\end{figure}

For out-of-plane wavevectors,
the magnon dispersion could be mapped out over the entire magnetic
Brillouin zone. A set of constant-{\bf Q} scans is shown in Fig.
3.  The curves show results of a convolution of the full 3D magnon
dispersion scattering with the instrumental resolution, assuming a
Lorentzian lineshape. Consideration of the fully convoluted
dispersion was necessary for accurate determination of the peak
positions because of the steep dispersion of the magnon branches
in the orthogonal $(hh0)$ direction.  In data taken at 50 K ($\sim
2.5 \rm T_{N}$), the magnons are no longer present.

\begin{figure}
\includegraphics[width=1.00\linewidth]{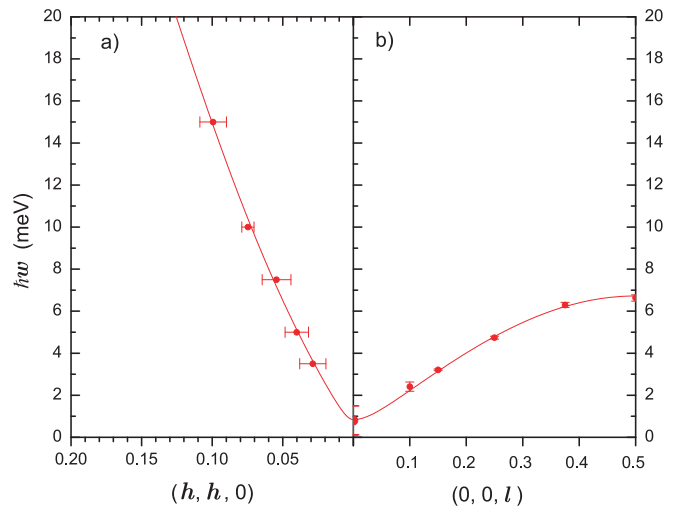}
\caption{a) Spin-wave dispersion along $(hh0)$; the peak positions
are from Fig. 2.  b) Dispersion along $(00l)$, with peak positions
from Fig. 3.  All data were taken at 1.6 K.  The dispersion curves
are the result of a fit to the model described in the text.}
\end{figure}

The dispersion data in both directions are summarized in Fig. 4.
Assuming nearest-neighbor interactions only, the spin Hamiltonian
of an A-type antiferromagnet with localized spins can be written
as
\begin{equation}
H={\rm J_{\parallel}}\!\!
\underset{in-plane}{\sum_{\left<i,j\right>} } \!{\bf S_{i}\cdot
S_{j}} + {\rm J_{\perp}}\! \underset{\perp
plane}{\sum_{\left<i,j\right>} }\!{\bf S_{i}\cdot S_{j}} - D\,
\underset{i}{\sum }\,\,{\rm S}^z_i\,,
\end{equation}
where ${\bf S_i}$ is the spin-1/2 operator for the magnetic ion at
lattice site $i$ and the coupling constants J$_{\parallel}$ and
J$_{\perp}$ characterize the exchange interactions within the
$ab$-plane and between adjacent planes, respectively.  The
anisotropy constant $D$, the sign of which alternates from layer
to layer, models the exchange anisotropy; it quantifies the
tendency of the spins to align along the {\it c}-axis. The magnon
dispersion can be calculated using the Holstein-Primakoff
formalism \cite{lovesey}; the resulting spin-wave dispersion
expression for ${\bf q}=(hkl)$ is
\begin{equation}
\hbar\omega\!=\!2{\rm S}\sqrt{\left\{\mathcal
J_{\perp}(0)\!-\!\left[\mathcal J_{\parallel}\!(0)\!-\!\mathcal
J_{\parallel}\!({\bf q})\right]\!+\!(D/2{\rm S})\right\}^{2}
\!\!-\!\left\{\mathcal J_{\perp}({\bf q})\right\}^{2}}\!,
\end{equation}
where $\mathcal J_{\parallel}\!({\bf q})\!=\!2 {\rm J_{\parallel}}
\left[\cos(2\pi h) +\cos(2\pi k) + \cos(2\pi (h+k))\right]$ and
$\mathcal J_{\perp}({\bf q})=2 {\rm J_{\perp}} \cos(\pi l)$. We
fitted the dispersion data in the $(hh0)$- and $(00l)$-directions
simultaneously and extracted the following values: J$\rm
_{\parallel}\!=\!-\,4.5\pm0.3$ meV, J$\rm _{\perp}\!=\! 3.3\pm0.3$
meV, and $|D|\!=\!0.05\pm0.05$ meV. Given the error in the peak
position associated with the energy scan through the zone center
at (003), and the large error in the fitted value of the
anisotropy parameter, the question of whether there is an
excitation gap at the AF zone center cannot be answered
definitively by our data, and awaits further experiments using
cold neutrons. Due to the dilution of spin-1/2 sites assumed for
$x=0.82$, the above calculation is only quantitatively accurate if
the material is phase-separated into magnetically ordered regions
with a dense network of spin-1/2 sites and nonmagnetic regions.
Another scenario, in which the spin-1/2 sites form an ordered
superlattice, is discussed below.  However, if the charge is
uniformly distributed in the CoO$_2$ layers or exhibits a small
modulation in density, the calculation should be repeated using an
itinerant model.

Several aspects of our data are surprising. First, the spin-wave
dispersion along the $ab$-plane is considerably steeper than that
along the {\it c}-direction, but the magnon bandwidth is
proportional to the number of nearest neighbors, which is six
within the $ab$-plane, and only two along $c$. Hence, J$\rm
_{\parallel}$ and J$\rm _{\perp}$ are in fact comparable in
magnitude. This relative isotropy is surprising in light of the
two-dimensionality of the Na$_{x}$CoO$_{2}$ crystal structure.  In
other layered magnets with comparable bond length anisotropies,
such as $\rm YBa_{2}Cu_{3}O_{6}$, the magnitudes of the in-plane
and out-of-plane exchange parameters differ by orders of magnitude
\cite{tranquada}.  It is also surprising that the Curie-Weiss
temperature inferred from the fitted exchange parameters in the
context of a local-moment picture, $(6 \rm J _{\parallel}+2J\rm
_{\perp})/k_B$, is positive, whereas that extracted from the
magnetic susceptibility measurements is negative.

The microscopic origin of these findings should be addressed by
theory. Since an unusually strong $c$-axis superexchange coupling
through Na or an unusually weak nearest-neighbor in-plane exchange
coupling would explain the isotropy of the spin wave dispersions,
a computation of these quantities is particularly desirable. A
recent theoretical study has pointed out the relevance of
longer-range exchange interactions along $c$ \cite{johannes}.
Another possible origin of our observation is an in-plane
charge-ordered superstructure recently suggested pursuant to NMR
\cite{ning,mukhamedshin} and optical conductivity \cite{christian}
experiments.  The
distribution of cobalt valence states has thus far not been
elucidated directly.  If all of the Co ions are in low-spin
local-moment states (that is, S = 1/2 for Co$^{4+}$ and S = 0 for
Co$^{3+}$), then for $x$ = 0.82 the spin lattice is dilute, with
only 18\% of the Co sites occupied by S = 1/2 spins. One possible
arrangement of these spins, achievable exactly for $x$ = 0.75, is
that of a triangular lattice with lattice constant $2a$. Since
this is comparable to the out-of-plane Co-Co distance, isotropic
spin wave dispersions would be a natural consequence. Such a spin
ordering could be inferred from more complete measurements of the
$ab$-plane magnon dispersion. A scenario in which the Co$^{3+}$
ions in the superstructure are in an intermediate-spin state with
S = 1 and interact antiferromagnetically \cite{christian} could
help explain why the bulk DC susceptibility in
Na$_{0.82}$CoO$_{2}$ is AF, despite the fact that $6 \rm J
_{\parallel}+2J\rm _{\perp}>0$.

If the cobalt valence is in fact distributed relatively uniformly,
an itinerant-electron picture should be considered instead. In
this case, the magnetic ordering would correspond to a
spin-density wave.


Another interesting aspect of our results is the q-width of the
$(hh0)$ magnons, which increases with increasing $|{\bf q}|$ (Fig.
2). The large peak widths may reflect short-ranged magnetic
correlations within the $ab$-planes.
An alternate possibility is Landau damping by charged
quasiparticles. In a charge-ordered scenario with
antiferromagnetically interacting, intermediate-spin Co$^{3+}$
ions arranged on a geometrically frustrated Kagom\'{e} lattice
\cite{christian}, the broadening may also arise from an admixture
of excitations from the disordered array of Co$^{3+}$ spins.

In conclusion, our measurements demonstrate that
Na$_{0.82}$CoO$_{2}$ exhibits spin fluctuations characteristic of
low-temperature 3D AF ordering of the A type, with magnetic
exchange constants much less anisotropic than expected, given the
layered crystal structure. The antiferromagnetism coexists with
metallic conductivity. Unanswered questions remain regarding the
degree to which the spins have localized or itinerant character,
possible superstructure in the former case, and the microscopic
character of the exchange couplings.  Answers to these questions
may also shed light on the origin of the unusual thermal
properties of Na$_{x}$CoO$_{2}$ and on the superconducting state
in the hydrated analogue.

We thank C. Bernhard, A.T. Boothroyd, A. Ivanov, G. Khaliullin,
R.K. Kremer, J. Kulda, P. Lemmens, I.~Mazin, W. Pickett, and R.
Zeyher for helpful discussions, and E. Br\"ucher for technical
assistance.

\end{document}